\documentclass[twocolumn,showpacs,amssymb,aps,nofootinbib,floatfix]{revtex4}
\usepackage{epsfig}
\usepackage{graphicx}
\newcommand{\be}{\begin{eqnarray}}
\newcommand{\ee}{\end{eqnarray}}

\newcommand{\mat}[1]{ \vec{\vec{#1 }}_{i j}}

\begin{document} \hbadness=10000
\topmargin -0.8cm\oddsidemargin = -0.7cm\evensidemargin = -0.7cm
\title{Instability of Boost-invariant hydrodynamics with a QCD inspired bulk viscosity}
\author{Giorgio Torrieri$^{a,b}$, Igor Mishustin$^{b,c}$}
\affiliation{$\phantom{A}^a$ITP,
  J.W. Goethe Universit\"at, Frankfurt A.M., Germany,  
torrieri@fias.uni-frankfurt.de}
\affiliation{$\phantom{A}^b$FIAS,
  J.W. Goethe Universit\"at, Frankfurt A.M., Germany  
mishus@fias.uni-frankfurt.de}
\affiliation{$\phantom{A}^c$Kurchatov Institute, Russian Research 
Center, Moscow 
123182, Russia.}
\date{May 5, 2008}  
\begin{abstract}
We solve the relativistic Navier-Stokes equations with homogeneous boost-invariant boundary conditions, and perform a stability analysis of
the solution. We show that, if the bulk viscosity has a peak around $T_c$ as inferred from QCD-based arguments, the background solution "freezes" at $T_c$ to a
nearly constant temperature state. This state is however highly unstable
with respect to certain inhomogeneous modes. Calculations show that these
modes have enough time to blow up and tear the system into droplets. We
conjecture that this is how freeze-out occurs in the QGP created in heavy ion collisions, and perhaps similar transitions in the early universe.
\end{abstract}
\maketitle
%%%%%%%%%%%%%%%%%%%%%%%%%%%%%%%
The system produced at RHIC \cite{announcement} is believed to be a good liquid in the early stages of its evolution.  At a later stage this liquid  transforms into a gas of particles which interact more weakly and eventually decouple.

The transition from a strongly interacting liquid into particles is however still not understood, both on the fundamental and phenomenological level.
On a conceptual level, no adequate description exists
of how the mean-free path goes from zero (``nearly ideal liquid'') to a distance comparable to system size (``transport'' of particles).

On a phenomenological level, hydrodynamics fails to reproduce particle 
interferometry data \cite{hydroheinz} unless the system decouples at an 
unrealistically high temperature \cite{hbtsolved} ($\sim T_c$, the critical 
temperature for the QCD phase transition) .   Moreover, attempts to 
use a hadronic transport model as an afterburner to hydrodynamics 
(conceptually considered to be the ``next best'' approximation) has 
failed to improve the model-data agreement 
\cite{hirano,shuryak,bass}.

Recent attempts to reconcile hydrodynamics with interferometric data \cite{pratt,ourbulk} have focused on a supposed sharp increase, and perhaps divergence, of bulk viscosity near $T_c$. This behavior of bulk viscosity  has now been inferred from a variety of
arguments\cite{visc1,visc3,visc4}.
In this work, we combine the recently acquired understanding of viscosity with 
an earlier study \cite{stability} to get a simple picture of how the peak in viscosity triggers freeze-out process.

% The conjectured peak in bulk viscosity causes the Boost-invariant solution to ``freeze'' at $T\sim T_c$.  This ``frozen'' system, however, becomes unstable against small perturbations.   Thus, it will be quickly torn apart into inhomogeneities.   These droplets, than could emit particles by evaporation, in a manner conjectured in \cite{ourbulk}.

The Navier-Stokes equations with Boost-invariant symmetry \cite{bjorken,background} can be rewritten \cite{stability} in terms of the Reynolds number $R$, the entropy $s$ the co-moving time $\tau$, the total number ($N$) of dimensions, and the dimensionality of the homogeneous expansion ($M$)
\begin{equation}
\tau^{-M} \frac{ d (\tau^M s)}{d\tau} = \frac{M s}{R\tau}
\end{equation}
For example, $M=1$ $N=3$ corresponds to the case studied in \cite{bjorken}, $M=N=3$ to the ``Krakow model'' \cite{krakow} or to a Friedman-like solution in flat space.

The Reynolds number is a function of temperature $T$,bulk and shear viscosity $\zeta$ and $\eta$ and entropy $s$
\begin{equation}
R^{-1}= \frac{2(1-M/N)\eta+M\zeta}{T s \tau}
\end{equation}
Given expressions for $s,\eta,\zeta$  in terms of $T$, this set of equations becomes closed and solvable.

For the equation of state, we use the parameterization of the speed of sound $c_s^2$ given in \cite{choj} (We have checked that our results do not vary qualitatively if the ideal EoS is used).  We use the ``sCFT limit'' \cite{adscft} for $\eta$ ($\eta=s/4 \pi$) and a parametrization of \cite{visc1,visc3} for $\zeta$
\begin{equation}
\zeta = s \left(  z_{pQCD} + \frac{z_0}{\sqrt{2 \pi}\sigma} \exp \left[- \frac{t^2}{2 \sigma^2}\right]  \right)
\end{equation} 
where $t=T-T_c$ and $\sigma=0.01 T_c$ and $z_{pQCD} \sim 10^{-3}$ \cite{viscT}.  
At $T>T_c$ (through not at the hadronic regime, $T<T_c$), this ansatz provides a reasonable fit to \cite{visc1}.    

We follow the stability analysis performed in \cite{stability}.  The amplitude of a generic perturbation to the 1D Boost-invariant system is a vector $\vec{x}$ in the two dimensional space of entropy perturbations and flow (rapidity $y$) perturbations
\begin{eqnarray}
 x_1=\frac{\delta s}{s} \phantom{A},\phantom{A} x_2 = y-y_{spacetime}
\end{eqnarray}
and its dependence on rapidity can be decomposed into Fourier components in rapidity space of wavenumber $k$ 
\begin{equation}
\vec{x}(y) = \sum_k \vec{x}(k) e^{ i k y}
\end{equation}
The equation of motion for $\vec{x}$ will then be given by
\begin{equation}
\label{perturbeq}
\tau \frac{\partial}{\partial \tau}  \left(  \begin{array}{c} x_1 \\i x_2  \end{array} \right)
 = \left( \begin{array}{cc}
A_{11} & A_{12}\\
A_{21} & A_{22}
\end{array} \right)
 \left(  \begin{array}{c} x_1 \\i x_2  \end{array} \right)
\end{equation}
where $\mat{ A}$ is a real matrix function of $s,R$ and $k$.  We refer the reader to Eq. 4.23-4.26 of \cite{stability} for the full form of $\mat{ A}$.  The evolution of perturbations can be understood through the behavior of the modulus of $\vec{x}$, $X = \vec{x}^T \vec{x}$
\begin{eqnarray}
\tau \frac{\partial}{\partial \tau} X = \vec{x}^T \mat{ M} \vec{x} 
\end{eqnarray}
Since $\mat{ M}$ is real and symmetric, it will always have two real Eigenvalues, $\lambda_{max}$ and $\lambda_{min}$ (corresponding Eigenvectors $\vec{x}_{max,min}$), as well as orthogonal matrices $\mat{B} $ diagonalizing it.    Defining $\vec{y}_i = \mat{B}^{-1} \vec{x}_j$we see that
\begin{equation}
\lambda_{min} y_{min}^2  <\tau \frac{\partial X}{\partial \tau} < \lambda_{max} y_{max}^2
\end{equation}
Thus, if $\lambda_{min}>0$, the system is inherently unstable, since perturbation in any direction will produce a positive growth rate.  If 
 $\lambda_{max}<0$, on the other hand, the system will be stable against all perturbations.

If $\lambda_{min}<0$ and $\lambda_{max}>0$ some modes will be stable and some will be unstable.   
In the latter case, the time dependence of $\mat{A}$ will in general continuously rotate $\vec{x}_{min}$ and  $\vec{x}_{max}$ in time: While an initial perturbation in the $\vec{x}_{max}$ direction grows as a power-law with $\tau/\tau_{ini}$, where $\tau_{ini}$ is the starting time of the perturbation
\begin{equation}
\vec{x}(\tau) \sim \left( \frac{\tau}{\tau_{ini}} \right)^{\lambda_{max}} \vec{x}_{max}(\tau_{ini})
\end{equation}
 its growth might well be stopped by a change of direction of the eigenvalues $\mat{A}(\tau)$: Under a general evolution of $\mat{A}$, $\vec{x}_{max}(\tau_{ini})$ might be in the direction of  $\vec{x}_{min}(\tau>\tau_{ini})$ a short time later.
Solving Eq. \ref{perturbeq} will take this effect into account.

%%%%%%%%%%%%%%%%%%%%%%%%%%%%%%
\begin{figure*}[t]
%\epsfig{width=8cm,clip=1,figure=plot_stat_z0.eps}
%\hspace{1cm}
%\vspace{1cm}
%\epsfig{width=8cm,clip=1,figure=plot_stat_z0.1.eps}
%\vspace{1cm}
%\epsfig{width=8cm,clip=1,figure=plot_stat_z1.eps}
%\hspace{1cm}
%\vspace{1cm}
%\epsfig{width=8cm,clip=1,figure=plot_stat_z10.eps}
\epsfig{width=18cm,clip=1,figure=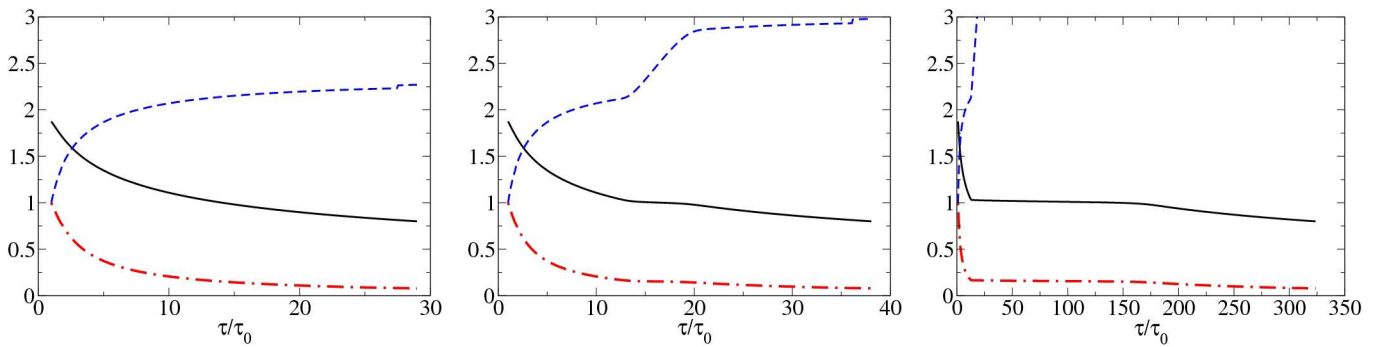}
\caption{\label{statplot}(Color online) Thermal parameters as a function of time for $z_0=0,0.1,1 T_c$ Solid line shows $T/T_c$. Dot-dashed  shows entropy density $dS/dy$ normalized to initial value, and Dashed the integrated entropy $S$ from an initial unit of rapidity, again normalized to the initial value. }
\end{figure*}
%%%%%%%%%%%%%%%%%%%%%%%%%%%%%%%

In what follows, we use $z_0=0,0.1,1$  $T_c$ 
 and evolve the system from an initial temperature $T=0.3$ 
GeV and comoving time 0.6 $fm$. Note that the qualitative features of our study are independent of the details of the evolution before $T_c$, such as 
the initial temperature and timescale. 

Fig. \ref{statplot} shows the temperature, entropy density and total entropy in the central rapidity unit as a function of time.   As can be seen, as soon as $z_0$ becomes non-negligible (i.e. viscous forces dominate around $T_c$), the kinematic evolution of the system ``freezes''.
The system then stays at nearly constant temperature, through it keeps producing more and more entropy at the expense of advective energy.   

At first sight, large values of $z_0$ are excluded by HBT data and multiplicity measurements \cite{etele}.   However, we will show that this long phase is unstable against small perturbations.   Thus, its further evolution will not be given by the background solution, but by a rapid formation of local inhomogeneities.

%%%%%%%%%%%%%%%%%%%%%%%%%%%%%%
\begin{figure*}[t]
%\epsfig{width=8cm,clip=1,figure=plot_lamb_z0.eps}
%\hspace{1cm}
%\vspace{1cm}
%\epsfig{width=8cm,clip=1,figure=plot_lamb_z0.1.eps}
%\vspace{1cm}
%\epsfig{width=8cm,clip=1,figure=plot_lamb_z1.eps}
%\hspace{1cm}
%\vspace{1cm}
%\epsfig{width=8cm,clip=1,figure=plot_lamb_z10.eps}
\epsfig{width=18cm,clip=1,figure=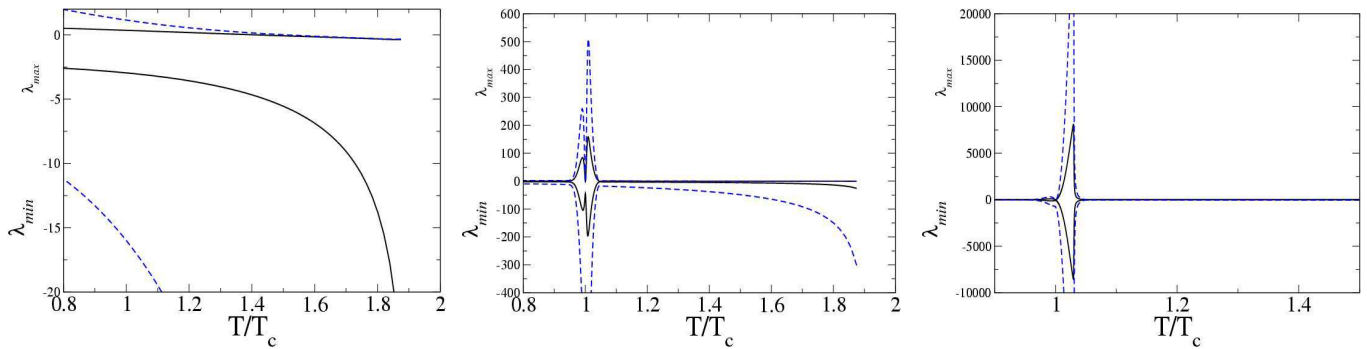}
\caption{\label{lambplot}(Color online) $\lambda_{min}$ and $\lambda_{max}$ as a function of $T/T_c$, for $z_0=0,0.1,1 T_c$. Only mixed stability data-points are shown (at later times in the $z_0=0$ case, evolution becomes stable). 
Solid black corresponds to $k=2$,dashed blue to $k=8$}
\end{figure*}
%%%%%%%%%%%%%%%%%%%%%%%%%%%%%%%
Fig. \ref{lambplot} shows the $\lambda_{min}$ and $\lambda_{max}$ Eigenvalues corresponding to representative $k=2,8$ (other values of $k$ were checked not to vary significantly wrt those presented here).   As can be seen, the peak in bulk viscosity forces the growing/damping rates to increase rapidly.   Thus, any initial perturbation in the unstable direction will rapidly grow to a value comparable with the background, unless the system's evolution will stop the growth by rotating the direction of the unstable modes.   

\begin{figure*}[t]
%\epsfig{width=8cm,clip=1,figure=plot_eig_z0.eps}
%\hspace{1cm}
%\vspace{1cm}
%\epsfig{width=8cm,clip=1,figure=plot_eig_z0.1.eps}
%\vspace{1cm}
%\epsfig{width=8cm,clip=1,figure=plot_eig_z1.eps}
%\hspace{1cm}
%\vspace{1cm}
%\epsfig{width=8cm,clip=1,figure=plot_eig_z10.eps}
\epsfig{width=18cm,clip=1,figure=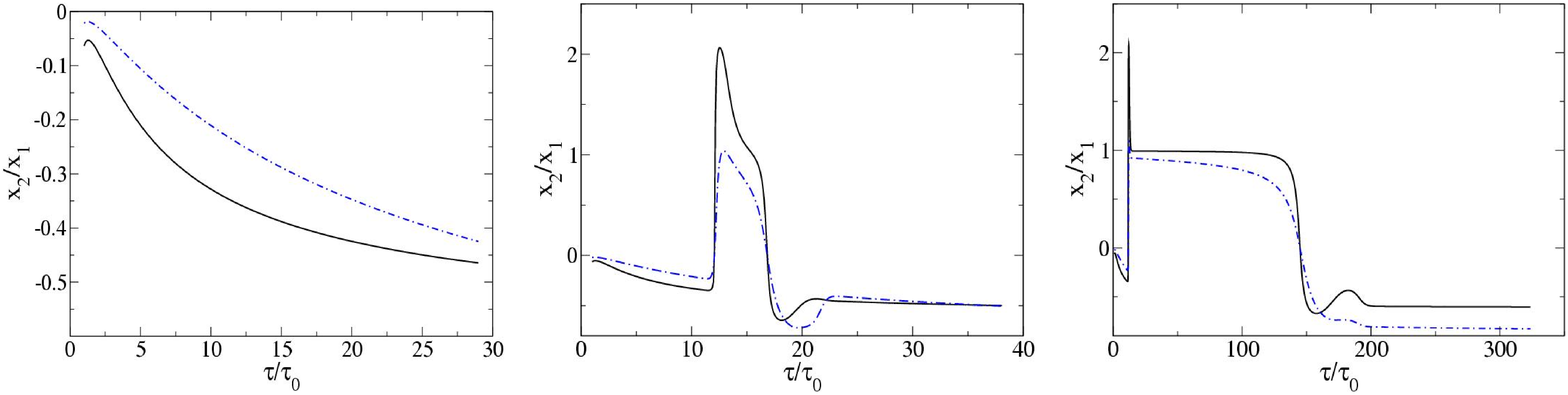}
\caption{\label{eigplot}(Color online) Direction of the unstable Eigenvector $\vec{x}_{max}$ as a function of time  for $z_0=0,0.1,1 T_c$. Solid black corresponds to $k=2$,dashed blue to $k=8$  }
\end{figure*}

Fig. \ref{eigplot} examines weather this occurs for larger values of $z_0$.   If the peak of viscosity is negligible, the unstable eigenvector keeps rotating throughout the evolution of the system.   Thus, even an unstable mode's growth will very quickly stop growing since the dynamics will turn it into a damped mode.   
When $z_0$ dominates, however, something every interesting happens:
At the time when $T$ approaches $T_c$, the direction of the unstable modes experiences an abrupt rotation.     Then it stays constant throughout the time the system travels through the viscosity peak (this time increases strongly as $z_0$ increases), and gets rotated again as the peak is passed.  The reason for this behavior is clear:  Fig \ref{lambplot} shows that the peaks occur at $T \simeq T_c$, where the background of the system is dynamically ``frozen'' (all advective energy is turned into entropy by viscous processes to keep temperature nearly constant).  Thus, it is not surprising that the direction of the unstable Eigenvector also remains approximately constant.

%%%%%%%%%%%%%%%%%%%%%%%%%%%%%%
%%%%%%%%%%%%%%%%%%%%%%%%%%%%%%%

Thus, at large $z_0$ unstable modes have all the required time to grow, especially considering the growth rate, $\sim \lambda_{max}$ in Fig. \ref{lambplot}, becomes overwhelming compared to the background expansion rate.

\begin{figure*}[t]
%\epsfig{width=8cm,clip=1,figure=growfile_tau_z0.eps}
%\hspace{1cm}
%\vspace{1cm}
%\epsfig{width=8cm,clip=1,figure=growfile_tau_z0.1.eps}
%\vspace{1cm}
%\epsfig{width=8cm,clip=1,figure=growfile_tau_z1.eps}
%\hspace{1cm}
%\vspace{1cm}
%\epsfig{width=8cm,clip=1,figure=growfile_tau_z10.eps}
%------PDF
%\includegraphics[width=8cm]{growfile_tau_z0.jpg}
%\includegraphics[width=8cm]{growfile_tau_z0.1.jpg}
%\includegraphics[width=8cm]{growfile_tau_z1.jpg}
%\includegraphics[width=8cm]{growfile_tau_z10.jpg}
%\includegraphics[bb =  0 0 590 451]{growfile_plot.jpg}
%\includegraphics[width=14cm]{growfile_plot.gif}
\epsfig{width=18cm,clip=1,figure=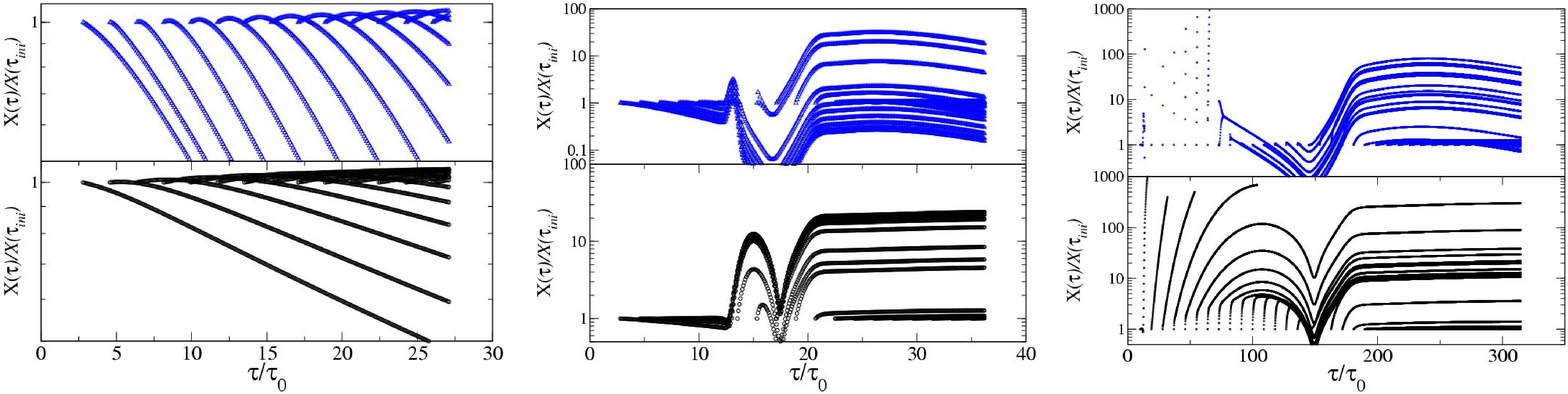}
\caption{\label{growplot}  Evolution of instabilities, starting at each timestep, for a $k=2$ perturbation  for, respectively $z_0=0,0.1,1 T_c$
Lower panel corresponds to $k=2$,upper panel to $k=8$}
\end{figure*}

Finally, Fig. \ref{growplot} shows the explicit solution of Eq. 
\ref{perturbeq}.   At each 
time-step, a perturbation is born in the unstable Eigenvector mode, and then evolved until the end of the evolution of the system. The plot shows $X(\tau)/X(\tau_{ini})$,  the ratio of strength of the perturbation to the initial strength as a function of time (Note that the starting value is always unity). A large $X(\tau)/X(\tau_{ini})$ does not mean the evolution equation is invalid: For any point in the graph, there will be a small enough perturbation amplitude that survives as a perturbation in the subsequent evolution.  The probability of a {\em larger} perturbation forming and significantly modifying the background, however, should grow strongly with $z_0$.   

Fig. \ref{growplot} makes it clear that any microscopic mechanism seeding instabilities at the scale $X \sim 10^{-1}$,uniformly distributed in $\vec{x}$ and at a rate of  $\sim fm^{-4}$ is likely to generate power-law growing instabilities  a few $fm$ after $T \sim T_c$.  These instabilities should reach $X \ge 1$, and hence play an important role in the subsequent evolution of the system, a few $fm$ after that.
%We have not as yet studied the mechanisms driving the formation of the instabilities, and hence can not speculate as to their initial distribution in $\vec{x}$ space.   However, if the distribution is any way uniform, their size $|\vec{x}|\sim 10^{-1}$, and the frequency of their seeding in spacetime is $\sim fm^{-4}$, it becomes clear that these instabilities will play an important role in the dynamics after a few femtometers of $T$ reaching $\sim T_c - \sigma$.

%%%%%%%%%%%%%%%%%%%%%%%%%%%%%%
%%%%%%%%%%%%%%%%%%%%%%%%%%%%%%%

Our model does not have the required scope to tell us what this role is likely to be.  Once grown, perturbations  will break local homogeneity and Boost-invariance, requiring full 3D hydro to be modeled properly.     

Transverse flow also has the potential to modify this scenario.   We note, however, that substituting 1D for homogeneous 3D Boost-invariant hydrodynamics \cite{stability}  does not quantitatively change our results.   As this ``3D Bjorken equation'' (corresponding to the hydro-inspired model examined in \cite{krakow}) is locally similar (up to rescaling) to the asymptotic behavior of realistic solutions describing transverse expansion \cite{csorgo}, we trust that the effect pointed out in this work is relevant for systems without transverse homogeneity.

While we leave the stability analysis in {\em second} order 
hydrodynamics (Israel-Stewart \cite{isr1,isr2,isr3} or its modern 
variations \cite{isrm1,isrm2,isrm3,isrm4,isrm5,isrm6}) to a future 
project, 
we do not expect it to play a big role in {\em starting} the instabilities:
As argued in \cite{ourbulk}, as long as the system's expansion rate as the system approaches $T_c$ is smaller than the relaxation time $\tau_{\Pi}$
\begin{equation}
\tau_\Pi \frac{1}{\sigma}\frac{dT}{d\tau} < 1
\end{equation}
second order terms will not prevent viscous corrections to the pressure of the order of $\zeta/\tau$, but merely localize their propagation.
The (admittedly unreliable) estimates from strongly coupled CFT \cite{janik} suggest that this criterion is amply satisfied, especially considering that the
increase in viscosity causes the background solution to slow down over a timescale much bigger than $\tau_\Pi$.
2nd order hydrodynamics might however play a dominant role in stopping the instabilities once they grow to a value comparable to the background.

Viscosity and causality should prevent any subsequent reinteraction of the created inhomogeneities. It is highly likely, therefore, that instabilities evolve isolated fragments, moving away from each other with pre-existing longitudinal and transverse flow.  This will have the effect of quickly stopping the unrealistically high entropy production seen in Fig. \ref{statplot}.

  The small dependence on $k$ of the instabilities growth rate (Fig. \ref{growplot}) might cause the formed inhomogeneities to continue seeding smaller and smaller instabilities, until the size of the smallest inhomogeneity becomes $\sim \Lambda_{QCD}^{-1}$, the scale at which  local expansion will be slowed down by the viscous forces,  The temperature of the center of such a ``cluster'' should be $\sim T_c$.   It is therefore tempting to regard these clusters, once formed and decoupled, as the ``fireballs'' within the Hagedorn picture \cite{hagedorn}.

This scenario shares some phenomenological similarities
with \cite{mish1,majumder,rafelski,kapusta} while differing in the fundamental description: In \cite{mish1,majumder,rafelski,kapusta},The role of the instability generator is the first order coexistence phase, and the instabilities are defined by being in a phase different from the ``background''.
In the current scenario, the instabilities are generated purely through hydrodynamic evolution, no discontinuities in the Equation of state are needed, and the perturbations are purely hydrodynamic in origin.  
%The role played in \cite{mish1,majumder,rafelski,kapusta} by the bag constant in seeding and stabilizing the perturbations is played, in our model, by viscous forces.

%It should also be noted that a high bulk viscosity signals the abandonment of the system of 
%chemical equilibrium \cite{murongachem}.  Thus, although this picture is physically
%different from \cite{rafelski}, it shares some phenomenological features without the
%theoretical assumptions: it does not require a ``sudden'' transition between regimes of
%differing entropy content to produce chemical non-equilibrium at $T_c$.

While this work focuses on heavy ion collisions, the dynamics examined here could play a role in cosmology, provided that quantum fields involved in cosmological phase transitions are approximately thermalized, and their viscosity depends on $T$ in a similar way to the one assumed here.
The general relativistic solution could than be locally similar enough to the (3D) Boost-invariant expansion (as in \cite{milne}) to allow for formation of local instabilities.
 These
could contribute to structure formation,  provide the loci where baryogenesis
would occur \cite{brandenberger}, or seed the formation of microscopic black holes.  The non-linear attractive behavior of gravity could enhance instability growth above hydrodynamic expectation.

In conclusion, we have shown that, when the conjectured rise of bulk viscosity at $T_c$ is inserted into the Navier-Stokes equations with Boost-invariant symmetry, the solution ``freezes'': While entropy continues to increase, temperature remains constant.
The background, however, becomes unstable against small perturbations, which then grow until boost-invariant hydrodynamics breaks down .
We hope that further work will clarify the subsequent evolution of these instabilities, and provide a quantitative link between this scenario and data.
%%%%%%%%%%%%%%%%%%%%%%%%%%%%%%%%%%%%%%%%%%%%%%%%%%%%%%%%%%%%%%%%%%%

%\acknowledgments 
\noindent GT thanks the Alexander von Humboldt Foundation and JW Goethe Universitat for their support.
IM acknowledges support provided by the DFG grant 436RUS 113/711/0-2 (Germany)
and grant NS-3004.2068.2 (Russia).
We thank P. Bozek for alerting us to \cite{stability}, R. Brandenberger 
for discussing the scenario presented here in a cosmological context, 
and J. Rafelski, L. Satarov and T. Koide for comments and suggestions.
%%%%%%%%%%%%%%%%%%%%%%%%%%%%%%%%%%%%%%%%%%%%%%%%%%%%%%%%%%%%%%%%%%5

\end{document}